\documentclass[fleqn,twoside]{article}
\usepackage{espcrc2}
\usepackage{times}
\usepackage{mathptm}

\usepackage{graphicx}
\usepackage[figuresright]{rotating}
\title{Final state interaction effects in neutrino-nucleus quasielastic 
scattering}

\author{C. Maieron\address{Istituto Nazionale di Fisica Nucleare, 
Sezione di Catania,
Via S. Sofia 64, I-95123 Catania, Italy },
        M.C. Mart\'{\i}nez\address[SEV]
        {Departamento de Fisica Atomica, Molecular y Nuclear, 
        Universidad de Sevilla, E-41080 Sevilla, Spain},
        J.A. Caballero\addressmark[SEV],
        and J.M. Ud\'{\i}as \address{Departamento de Fisica Atomica, 
        Molecular y Nuclear, 
        Universidad Complutense de Madrid, E-28040 Madrid, Spain}}
       
\begin{document}

\begin{abstract}
We consider the charged-current quasielastic scattering 
of muon neutrinos on an Oxygen 16 target,
described within a relativistic shell model and, 
for comparison, the relativistic Fermi
gas. 
Final state interactions are described in the 
distorted wave impulse approximation,
using both a relativistic mean field potential and 
a relativistic optical potential, with and without imaginary part.
We present results for inclusive cross sections at fixed neutrino 
energies in the
range $E_\nu =$ 200 MeV - 1 GeV, showing that final state
interaction effects can remain sizable even at large energies.

\vspace{1pc}
\end{abstract}

\maketitle

\section{INTRODUCTION AND FORMALISM}

\begin{figure}[h]
\begin{center}
\includegraphics[width=0.5\textwidth]{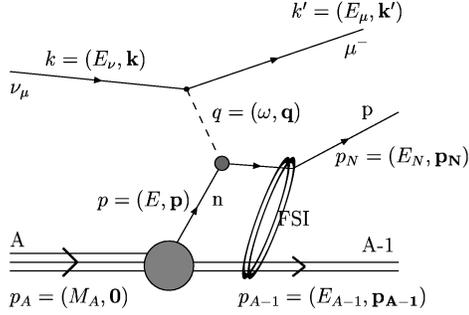}
\end{center}
\vskip -1 cm
\caption{
  Born approximation diagram for CC $\nu$--nucleus quasielastic
  scattering. The impulse approximation is assumed in the hadronic
  vertex and the possibility of final state interactions between the
  outgoing nucleon and the residual nuclear system is explicitly
  shown.
}
\label{fig:ia}
\end{figure}
\begin{figure}[h]
\vskip -1 cm
\includegraphics[width=0.55\textwidth]{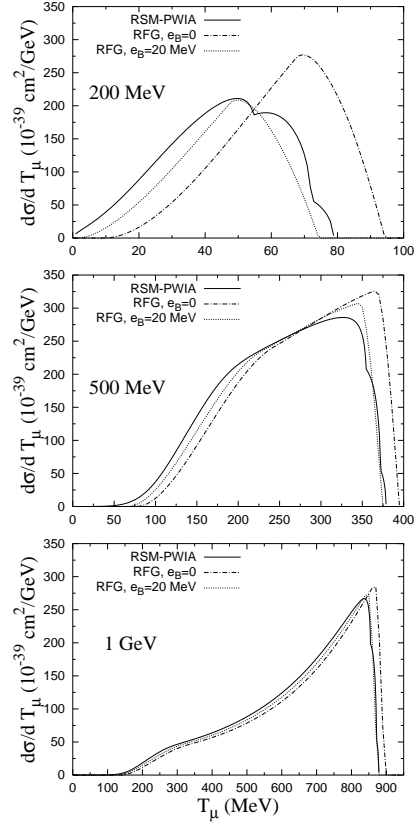}
\vskip -1 cm
\caption{
  Differential cross section $(d\sigma/dT_\mu)$
  versus the outgoing muon kinetic energy, for the quasielastic
  scattering of muon neutrinos on $^{16}O$ and incident neutrino 
  energy: $E_\nu=200$ MeV (upper panel),
  $500$ MeV (middle) and $1$ GeV (lower panel).}
\label{fig:sigma1}
\end{figure}
\begin{figure}[htb]
\vskip -1 cm
\includegraphics[width=0.55\textwidth]{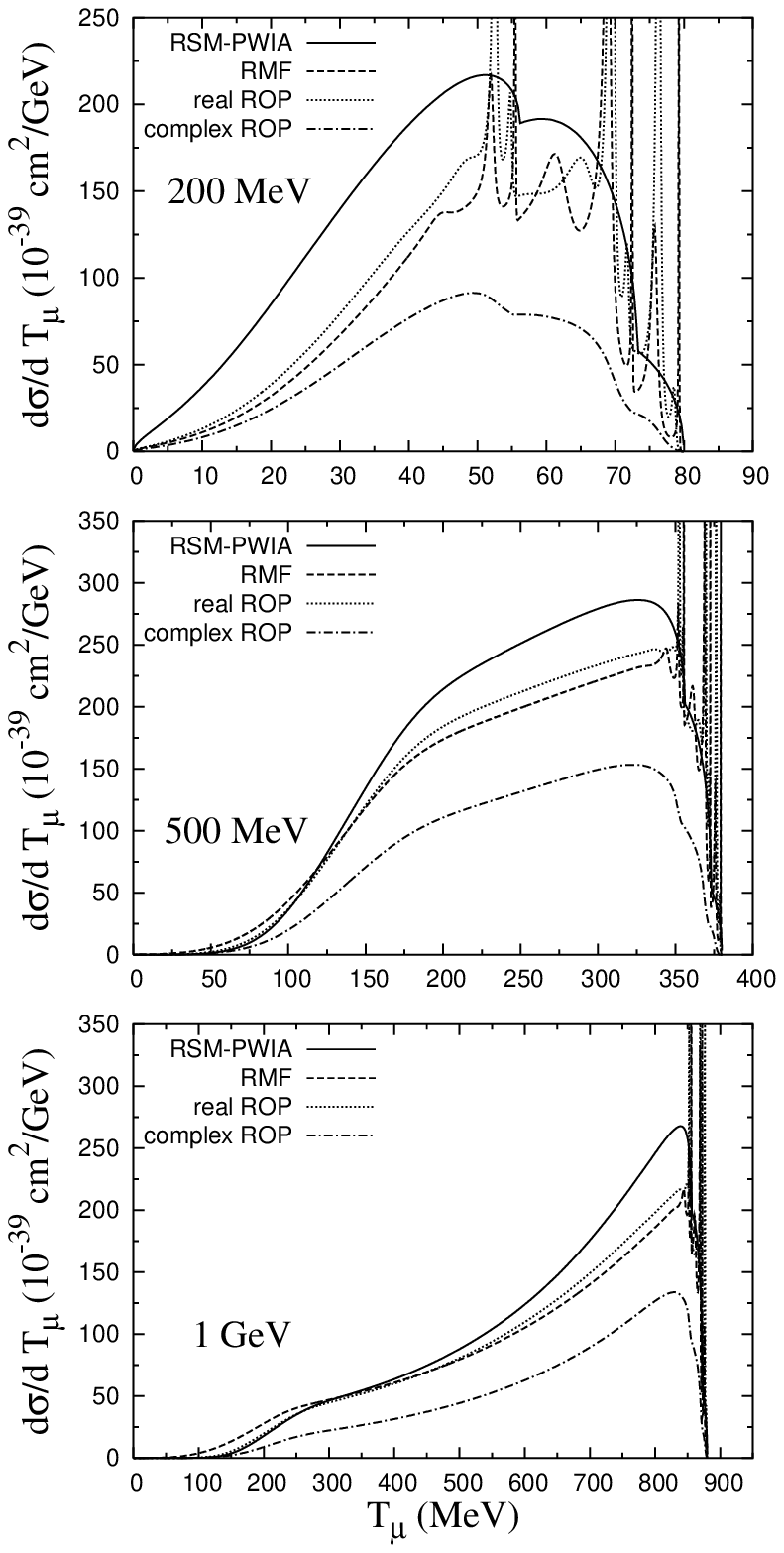}
\vskip - 1cm
\caption{Same as fig.~\ref{fig:sigma1}, but including
  FSI effects.  All curves are calculated using the RSM model,
  in PWIA (solid) and
  describing FSI within the RMF (dashed), real ROP (dotted) 
  and complex ROP (dot--dashed) approaches.
  }
\label{fig:sigma2}
\end{figure}
A precise and realistic description of neutrino-nucleus interaction
in the intermediate energy region is crucial for the interpretation
of experimental results used to determine neutrino properties.
At present most of the 
Monte Carlo codes \cite{MC} 
which have been developed to simulate the response
of the detectors in these experiments 
are based on the Fermi gas model. Therefore they 
take into account the Fermi motion of the nucleons inside the nucleus
and Pauli blocking effects, but they neglect several other
effects, which are known to be important from electron scattering 
experiments \cite{BenharGiusti}. 

In order to study the relevance of some of these
effects, in this contribution we compare relativistic Fermi gas 
(RFG) calculations
of neutrino-nucleus quasielastic scattering cross sections
\cite{Alb97}, with
results obtained within a relativistic shell model (RSM), 
including, in particular, final state interactions (FSI)
\cite{eep}.
More details about our calculations can be found in Ref.~\cite{noi}. 
FSI effects in $\nu$-nucleus scattering have also
been studied in Refs.~\cite{Co,Meucci}.

We consider the charged-current (CC) quasielastic scattering of muon
neutrinos on $^{16}O$ at fixed neutrino energies of
200, 500 and 1000 MeV, describing 
this process within the impulse approximation,
schematically represented in fig.~\ref{fig:ia}.

We thus assume
that the incident neutrino interacts with only one nucleon,
which is then emitted, while the remaining (A-1) nucleons in the
target are spectators, that the nuclear current is the
sum of single nucleon currents and that the states of the target and
residual nuclei are described by independent particle
model wave functions.  
We describe the ground state of  $^{16}O$ as a closed shell
configuration, the occupied shells being $s_{1/2}$, $p_{3/2}$ and
$p_{1/2}$. For the
removal of a nucleon from a closed shell of angular momentum $j$, the
cross section corresponding to the diagram in fig.~\ref{fig:ia} has
the following general form: 
\begin{eqnarray}
\lefteqn{\frac{d^6 \sigma}{d^3 k^\prime d^3 p_N}
=
\displaystyle{\int}\delta^4(q + p_A -p_{A-1} - p_N)
\frac{G_F^2}{(2\pi)^5}}\label{eq:dsigma}\\
\lefteqn{\times
\frac{(2 j + 1)}{8 E_{\nu}E_\mu}
\overline{\sum}
|
\bar{u}(k')
\gamma^\alpha
\left( 1 + \gamma_5 \right)
u(k)\,
J_\alpha({\bf q})|^2
d^3p_{A-1},}
\nonumber
\end{eqnarray}
where $G_F$ is the
Fermi constant, $\overline{\sum}$ indicates the average/sum over
the initial/final spins,
$\gamma_5 =
-i\gamma^0\gamma^1\gamma^2\gamma^3$ and Dirac spinors are normalized
  according to
$u(k)^\dagger u(k) = 2 k_0$.  We then sum over the occupied shells and
integrate over the emitted nucleon and over the direction of the
outgoing muon in order to get the inclusive cross sections
$d \sigma/d T_\mu$, and the integrated
inclusive cross section $\sigma = \displaystyle{\int\left(d \sigma/
dT_\mu\right)dT_\mu}$, $T_\mu$ being the outgoing lepton kinetic energy.

The main ingredient in eq.~(\ref{eq:dsigma}) is the single nucleon
current matrix element,
\begin{equation}
J_\alpha({\bf q}) =
\sqrt{V} \displaystyle{\int}
d^3 r e^{i{\bf q}\cdot {\bf r}} 
\, \overline{\psi}_{s_N}({\bf p}_N, {\bf r})
\hat{\Gamma}_\alpha \,
\psi_B^{jm}({\bf r})\;,
\label{eq:jsn}
\end{equation}
where $\psi_B^{jm}({\bf r})$ and $\psi_{s_N}({\bf p}_N,\,{\bf r})$ are
the wave functions for the initial (bound) nucleon and
for the emitted nucleon, respectively, and $\hat{\Gamma}_\alpha$ is the single
nucleon weak charged-current operator.  For the latter we assume the free,
on mass shell, nucleon expression:
\begin{eqnarray}
\hat{\Gamma}_\alpha &=& |V_{ud}| 
\left[F_V\gamma_\alpha + F_M 
\frac{i}{2 m_N}\sigma_{\alpha \beta} q^\beta
\right. \nonumber \\
&+& \left. F_A \gamma_\alpha \gamma_5 - F_P q_\alpha \gamma_5
\right] \, ,
\label{eq:gamma} 
\end{eqnarray}
where $F_{V,M}$ are the CC single nucleon Pauli and Dirac form
factors, $F_A$ and $F_P$ are the axial and the induced pseudoscalar
form factor, respectively, and $|V_{ud}|$ is
the $ud$ Cabibbo--Kobayashi--Maskawa matrix element.  
For $F_A$ we assume a dipole parameterization with cutoff mass
$M_A=1.026$ GeV. 

In eq.~(\ref{eq:jsn}) the bound nucleon wave functions
$\psi_B^{jm}({\bf r})$ are the self--consistent (Hartree) solutions of
a Dirac equation, derived, within a relativistic mean field approach,
from a Lagrangian containing $\sigma$, $\omega$ and $\rho$
mesons,which has been already successfully
used in the study of $(e,e'p)$ processes \cite{eep}.
For the single-particle binding energies of the
different shells we use the corresponding experimental values, which 
determine the threshold of the cross section for every shell.

Concerning the emitted nucleon, 
as a starting point we describe it as a plane wave,
thus neglecting its interaction with the
residual nucleus (plane wave impulse approximation - PWIA). 
Then, to make our description more realistic, 
we include the effects of FSI by using
distorted waves (distorted wave impulse approximation - DWIA).
For the latter, we make the following different choices.

{\it Complex ROP}: following previous studies
of exclusive electron scattering processes \cite{eep}, we employ distorted
waves which are obtained as solutions of a Dirac equation containing a
phenomenological relativistic optical potential (ROP). 
The ROP has a real part, which describes the rescattering of the
ejected nucleon and  an imaginary part, that accounts for the absorption
of it into unobserved channels. 

{\it Real ROP:}
since, contrary to the $(e,e'p)$ case, here we are 
considering inclusive processes, where all final channels
contribute, 
the presence of the
imaginary term in the optical potential leads to an
overestimation of FSI effects.
For this reason 
we also consider the potential obtained by setting the
imaginary part of the ROP to zero (a discussion
on the use of real optical potentials has been presented in 
Ref.~\cite{Meucci}).

{\it RMF:} finally, we employ 
wave functions which are obtained as the solutions in the continuum
of the same Dirac equation which is used to derive the bound
nucleon wave functions. We refer to this approach as relativistic mean field
(RMF) and consider it appropriate at low energy transfer. 

\section{RESULTS AND CONCLUSIONS}
Let us now present the results we obtain for neutrino-nucleus
quasielastic cross sections.
As a first step we neglect FSI effects and just compare RSM-PWIA
results with the corresponding curves obtained within the relativistic
Fermi gas, for which we use a Fermi momentum
$p_F=225$ MeV and binding energy $e_B=0$ or $20$ MeV.
This comparison is presented in fig.~\ref{fig:sigma1}, where
the differential cross sections $d\sigma/dT_\mu$ is plotted 
as a function of the outgoing muon kinetic energy.
We observe that the differences between the two models are 
quite large at low neutrino energy, but they practically disappear
at $E_\nu=1$ GeV.

The situation is different if we ``turn on'' FSI effects,
as illustrated
in fig.~\ref{fig:sigma2}, where the RSM-PWIA results of
 fig.~\ref{fig:sigma1} are compared with the DWIA approaches
previously outlined. 
 
We see that FSI effects
produce a reduction of the cross section, with respect to PWIA.
The RMF and real ROP curves are quite similar, showing a reduction
of about  
$30\div 40\%$ for $E_\nu = 200$ MeV and $20\%$ for the other energy
values. In the case of the complex ROP model, instead, the reduction 
of the cross section is rather large,
between  60\% ($E_\nu=200$ MeV) and 50\% (at higher energy),
due to the absorption introduced by the imaginary term. 

Let us note that the resonant structure appearing at high
$T_\mu$ is due to the
use of real potentials (RMF and real ROP) for describing the final nucleon
state. Here we only include single-particle excitations 
within a mean field picture, while including residual interactions would make
the width and number of resonances to be considerably larger.

\begin{figure}[t] 
\vskip -0.2 cm
\hskip -0.2 cm
\includegraphics[width=0.5\textwidth]{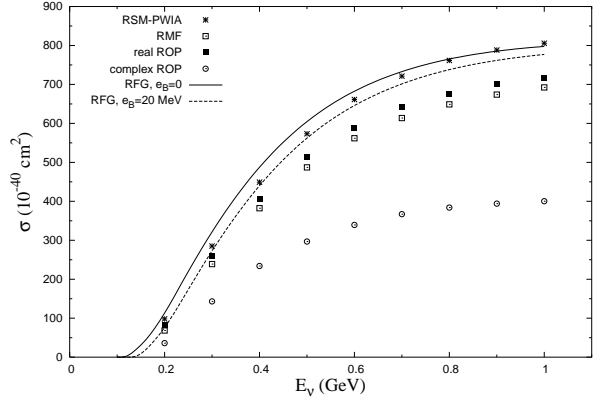} 
\vskip -0.5 cm
\caption{Integrated cross section $\sigma$ for the quasielastic
scattering of muon neutrinos on $^{16}O$ as a function of the incident
neutrino energy. The curves are calculated within the RFG model, while
the points correspond to RSM calculations without FSI (stars)
and with FSI effects taken into account within the RMF (empty squares),
real ROP (full squares) and complex ROP (circles) approaches.}
\label{fig:sigmatot} 
\end{figure}

Finally, the results of figures~\ref{fig:sigma1} and~\ref{fig:sigma2}
are summarized in fig.~\ref{fig:sigmatot}, 
where the cross section $\sigma$, integrated over the muon energy, is
plotted
as a function of the incident neutrino energy $E_\nu$.
Again we see that within
the PWIA the discrepancy between different nuclear models is relatively
small and decreases with increasing neutrino energy, while FSI
effects are still present even at large $E_\nu$. 
The (more reliable) results for
the RMF and real ROP approaches show a not too large, 
but still appreciable reduction ($\sim 15\%$ at $E_\nu=1$ GeV),
while, again, the
imaginary term in the complex ROP leads to a too large reduction ($\sim
50\%$) of the cross section.  

In conclusion, within the Impulse Approximation, we observe 
that when FSI effects are neglected,
the nuclear model dependence of the quasielastic $\nu$-nucleus cross section
is large at low neutrino energy but becomes negligible
at $E_\nu$ = 1 GeV.
FSI effects are also rather large at $E_\nu\simeq$ 200 MeV and decrease
with increasing neutrino energy, but, in the more realistic
real ROP approach, can still be as large as 15 \% at $E_\nu$ = 1 GeV
and thus must be carefully considered.

As a final remark, we notice that if we consider
Neutral Current quasielastic processes, where the outgoing
nucleon must be detected, then using the complex ROP
description of FSI would be more appropriate. Therefore in this case
rather large FSI effects are expected to be present even at relatively
high neutrino energy \cite{Alb97,MeucciNC}.

\end{document}